\documentclass[reprint,amssymb, amsmath, aip,cha,onecolumn]{revtex4-1}
\usepackage{graphicx}
\usepackage{color}
\usepackage{epsfig}
\usepackage{bm}%
\usepackage[colorlinks=true,linkcolor=blue]{hyperref}%
\expandafter\ifx\csname package@font\endcsname\relax\else
 \expandafter\expandafter
 \expandafter\usepackage
 \expandafter\expandafter
 \expandafter{\csname package@font\endcsname}%
\fi
\begin{document}
\title{Effect of magnetic field on the phase transition in a dusty plasma}%
\author{S. Jaiswal}%
\email{surabhijaiswal73@gmail.com}
\affiliation{Institut f{\"u}r Materialphysik im Weltraum, Deutsches Zentrum f{\"u}r Luft-und Raumfahrt (DLR), 82234 We{\ss}ling, Germany}%
\author{T. Hall}
\affiliation{Auburn University, 206 Allison Laboratory Auburn, AL 36849-5319}%
\author{S. LeBlanc}
\affiliation{Auburn University, 206 Allison Laboratory Auburn, AL 36849-5319}%
\author{R. Mukherjee}
\affiliation{Institute For Plasma Research, Bhat, Gandhinagar,Gujarat, India, 382428}%
\author{E. Thomas}
\affiliation{Auburn University, 206 Allison Laboratory Auburn, AL 36849-5319}%
\date{\today}
%**************************************************************
%#####################################################################################                  ABSTRACT
%************************************************************************************************
\begin{abstract}
The formation of self-consistent crystalline structure is a well-known phenomenon in complex plasmas. In most experiments the pressure and rf power are the main controlling parameters in determining the phase of the system. We have studied the effect of externally applied magnetic field on the configuration of plasma crystals, suspended in the sheath of a radio-frequency discharge using the Magnetized Dusty Plasma Experiment (MDPX) device. Experiments are performed at a fixed pressure and rf power where a crystalline structure is formed within a confining ring. The magnetic field is then increased from 0 to  1.28 T. We report on the breakdown of the crystalline structure with increasing magnetic field. The magnetic field affects the dynamics of the plasma particles and first leads to a rotation of the crystal. At higher magnetic field, there is a radial variation (shear) in the angular velocity of the moving particles which we believe leads to the melting of the crystal. This melting is confirmed by evaluating the variation of the pair correlation function as a function of magnetic field.
%%%%%%%%%
\end{abstract}
%%%%%%%%%%%%%%%%%%%%%%%%%%%%%%%%
\maketitle
%%%%%%%%%%%%%%%%%%%%%%%%%%%%%%%%%%%%%%%%%%%%%%%%%%%%%%%%%%%%%%%%%%%%%%%%%%%%%%%%%%%%%       INTRODUCTION
%%%%%%%%%%%%%%%%%%%%%%%%%%%%%%%%%%%%%%%%%%%%%%%%%%%%%%%%%%
\section{Introduction}\label{sec:intro}
\lq\lq Dusty" or \lq\lq complex" plasmas consist of the usual combination of electrons, ions, and neutral atoms with the addition of charged, dust particulates of size ranging from tens of nanometers to tens of micrometers. In the plasma environment, these heavy dust particles can acquire a negative charge of the order of Z$_d \sim 10^4-10^5$ elementary charges due to collection of more electrons than ions which adds much richness to the collective dynamics of the system. Dusty plasmas occur in a variety of natural situations such as planetary rings, comet tails, interplanetary media, and interstellar clouds \cite{goertz, nakamura_book}. Although the dust particles are highly charged, their mass is much larger than that of the electrons or ions, m$_d \sim 10^{-15}$ to $10^{-13}$ kg. As a result, the dust particles have a much smaller charge to mass ratio than electrons and ions, which makes the time scales of dust dynamics comparatively much longer. Additionally, the bigger sized and highly massive particles possess a very low thermal velocity, hence allow them to be visualized using laser illumination and high-speed cameras. Thus dusty plasma offers an excellent medium for studying various phenomena at single particle and fluid level with remarkable temporal and spatial resolutions.\par
In a typical laboratory experiment, the dust cloud is levitated near the sheath boundary by balancing the electrostatic force due to sheath electric field and the gravity. The highly charged particle interact each other via a strong electrostatic potential that may exceed the thermal energy and plasma become strongly coupled. The strength of coupling can be determined by a coupling parameter $\Gamma$ which is the ratio of interparticle coulomb potential energy to the dust thermal energy $(\Gamma =   Q^2/(4\pi \epsilon_0 aK_B T_d ))$. %In a number of theoretical and experimental studies it has been found that 
Several authors previously revealed that above a critical value of coupling parameter $\Gamma_c=171$ \cite{Ikezi, thomas_1994}, the dust can freeze into perfect crystal whereas in the regime of $1\ll \Gamma < \Gamma_c$, the system exhibits a strongly coupled fluid state. Another parameter is coulomb screening constant ($\kappa$), which is the ratio of interparticle spacing and the screening length. The $\kappa$ sometimes used as effective Coulomb coupling parameter \cite{kalman} $\Gamma_{eff}$, which is diminished by effect of shielding. Because it is possible to experimentally control the Coulomb coupling parameter, dusty plasma have been used to study the phenomenon of phase transition and transport properties of strongly coupled system.\par
A large number of experimental studies in the past have been devoted to the formation of dust crystal and its melting dynamics in a rf plasma. Some of the pioneering experiments in this area have been reported by Chu \textit{et al.}\cite{chu}, H. Thomas \textit{et al.}\cite{thomas_1994}, and Hayashi \textit{et al.}\cite{hayashi}. Chu \textit{et al.}\cite{chu} demonstrated the formation of coulomb crystal and liquid in a rf produced strongly coupled dusty plasma for a 10 micron SiO$_2$ particles. They  observed the hexagonal, fcc and bcc crystal structures and solids with coexisting different crystal structures with propagating boundaries at lower rf powers and identified the transition to the more disordered liquid state by increasing the rf power. Hayashi \textit{et al.}\cite{hayashi} demonstrated the formation of coulomb crystal as a result of growth of carbon particles in a methane plasma. They have shown the dependency of phase transition phenomenon over the particles diameter and Wigner-Seitz radius of the structure. H. Thomas \textit{et al.} \cite{thomas_1994}  investigated the plasma crystal and its phase transition by changing the various controlling parameter including plasma density, temperature, neutral gas, and particle size. Later, various theoretical and experimental research on the dust crystal and phase transition phenomena have been reported by several authors \cite{chu_lin, melzer_1994, G_morfill_jvst, Pieper_goree}. In all these investigations it has been concluded that the structural configuration of particle cloud is mainly controlled by the pressure, rf power and micro-particle size.\par 
In addition to the plasma crystal studies, over the last decade, a series of research focusing the effect of externally applied magnetic field over the particle dynamics have been reported. It has been well-documented that dusty plasma phenomena become even more complicated in presence of external magnetic field. Yaroshenko \textit{et al.} \cite{yaroshenko} studied the  various mutual dust-dust interactions in complex plasmas, including the forces due to induced magnetic and electric moments of the grains. Konopka \textit{et al.}\cite{konopka_2000} have observed the rotation of plasma crystal under the influence of constant vertical magnetic (0.014 T) at various discharge conditions in a rf plasma. The estimated shear stress in their experiment has been used to calculate the shear elastic modulus of the dust crystal. Plasma rotation due to azimuthal E$\times$B ion drift in a dc glow discharge plasma has been observed by Uchida \textit{et al.} \cite{uchida, sato} while the deflection of single-particle trajectory in a electron cyclotron resonance plasma has been examined by Nonumura \textit{et al.} \cite{nonumura_1997}. \par
%In a last couple of years, the phenomenon of Phase transition has been theoretically investigated by varying the magnetic field strength. In a paper by Baruah \et al 
Some recent theoretical works have shown that the phase transition phenomenon can also be controlled by varying magnetic field strength \cite{shukla_dust_crystal, avinash_crystal, baruah_crystal, begum_crystal}. Presence of magnetic field alters the internal energy of the particulates in plasma and hence affects the points of phase transition from solid to liquid state. %The effective interaction potential among the dust grains is modified in the presence of magnetic field.
%In a M.D. simulation by 
Baruah \textit{et al.} \cite{baruah_crystal} have adopted that modified interaction potential into their molecular dynamic simulation and found that the crystalline behavior changes by changing the magnetic field strength and system turns into fluid or gaseous states above a very high magnetic field (above 1 T). However, the direct effect of magnetic field on the dust grain dynamics has not been taken into account. Mahmuda \textit{et al.} \cite{begum_crystal} have reported that the self-diffusion is a criterion for melting of dust crystal in the presence of magnetic field. A detailed investigation of crystalline behavior and phase transition of dusty plasma in presence of magnetic field is important for the point of understanding of fundamental physics as well as for production and control of the properties of dust crystal. To understand these effects in depth, a dedicated experimental verification is indeed needed. \par 
In this paper, we have experimentally investigated the effect of externally applied magnetic field on the configuration of plasma crystals in the sheath of an rf discharge argon plasma. The experiment has been performed in the Magnetized Dusty Plasma Experiment(MDPX) device with mono-disperse melamine formaldehyde particles. The plasma crystal has been formed within a circular ring at a particular discharge condition and the melting of the crystal has been observed at higher magnetic field (at 1 T). The melting of the crystals is confirmed by plotting the pair correlation function and cause of melting is qualitatively explained by radial variation of angular velocity measurement of the moving particles.\par
The paper is organized as follows.  In the next section (Sec.~\ref{sec:setup}), we present the experimental set-up in detail. In Sec.~\ref{sec:results}, we discuss the experimental results on the formation of dust crystal and its melting dynamics.  A brief concluding remark is made in Sec.~\ref{sec:conclusion}.
 %%%%%%%%%%%%%%%%%%%%%%%%%%%%%%%%%%%%%%%%%%%%%%%%%%%%%%%%%%%%%%%%%%%%%%%%%%%%%%%%%%%%%%%%%%%%%
 %&&&&&&&&&&&&&&&&&&&&&&&&&&&&&&&&&&&&&&&&&&&&&&&&&&&&&&&&&&&&&&&&&&&&&&&&&&&&&&&&&&&&&&&&&&&&&&&&&&&&&&&&&&&&&&&
 %***********************************************************************************************************************
\section{Experimental Setup}\label{sec:setup}
The Magnetized Dusty Plasma Experiment (MDPX) device is a recently commissioned, multi-user, high magnetic field experimental platform that is operating at Auburn University. The MDPX device consists of two main components:  the superconducting magnet and the primary plasma chamber.  The hardware components of the MDPX device are described extensively in previous papers \cite{ed_1, ed_2, ed_3}.  For the studies described in this paper, the MDPX device is operated in its vertical configuration with the magnetic field aligned parallel to gravity. All four superconducting coils are energized at the same current to produce a uniform magnetic field (with $\Delta B/B < 1 \%$) at the center of the experimental volume where the plasma and dusty plasma are generated.\par
An octagonal vacuum chamber with a 355 mm circular inner diameter is used as the plasma source for the MDPX device. The basic configuration for the plasma generation uses a pair of 342 mm diameter electrodes to produce a capacitively coupled, glow discharge argon plasma using a 13.56 MHz radio frequency source connected to a matching network. The lower electrode is powered while the upper electrode is electrically grounded.  The upper electrode has a 146 mm circular hole to allow viewing of the plasma and the dust particles from the top. The lower electrode has a 152 mm wide $\times$ 3.2 mm circular depression to aid in the confinement of the dust particles. For this experiment, in the center of the depression, a copper ring that has a 50 mm outer diameter, 41 mm inner diameter, and 1.5 mm thickness is placed to further restrict the size of the particle cloud formed by the suspended particles.\par
\begin{figure}[!hb]
\includegraphics[scale=0.5]{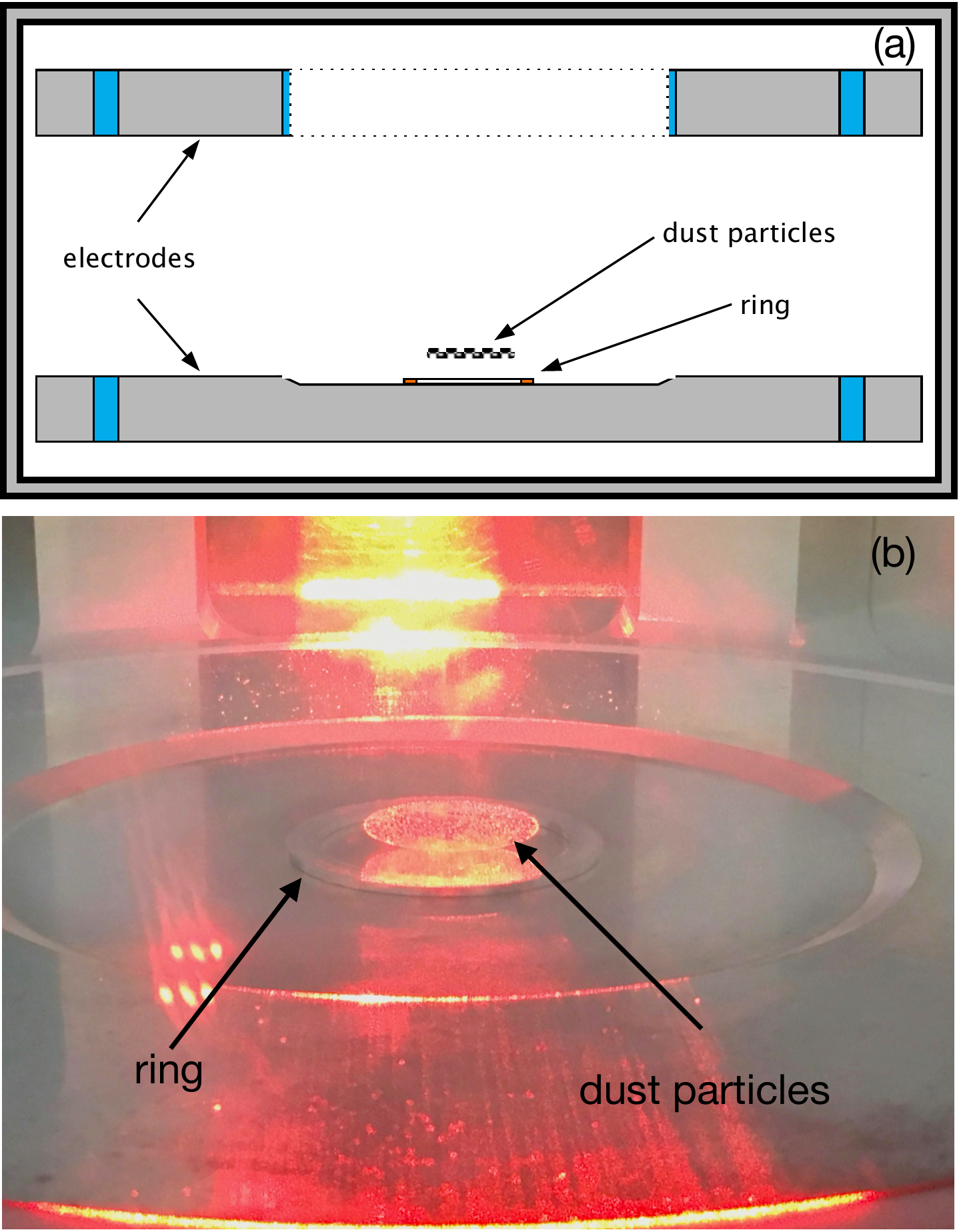}
\caption{\label{fig:setup} a) A schematic drawing of the interior of the MDPX plasma chamber and b) A photograph of a plasma crystal suspended in the plasma.}
\end{figure}

In this experiment, mono-disperse melamine formaldehyde particles of diameter $7.17\pm0.08$ $\mu$m are used as the dust component. The particles are introduced using a radial \lq\lq shaker" that is placed into the center of the plasma discharge and tapped to release a few hundred particles. The shaker is then pulled radially outward to avoid disturbing the plasma. The particles are illuminated using a 633 nm, red laser diode that is expanded into a thin, horizontally aligned sheet. The particles are viewed from the top port of the vacuum chamber using a USB3-based, 2048 by 2048 pixel, Ximea model xiQ camera that can be operated up to 90 frames per second (fps); although for this experiments frame rates of 10 to 60 fps were sufficient to perform these studies. Combinations of particle tracking velocimetry (PTV) and particle image velocimetry (PIV) techniques were used to measure the particle dynamics \cite{ed_4, ed_5, ed_6}. Fig.~\ref{fig:setup}(a) shows a schematic drawing of the interior arrangement of the plasma source and Fig.~\ref{fig:setup}(b) showing photograph of a plasma crystal that is suspended above the ring in the MDPX plasma chamber.\par
A Langmuir probe has been used to measure the plasma conditions for B = 0 T and without the presence of the dust particles which yields an electron temperature, T$_e \sim 2.5$ to 3.5 eV and a plasma density of n $\times~0.5$ to $3\times10^{15}$  m$^{-3}$.  It is noted that once the magnetic field is energized, the probe traces become severely distorted and are effectively unreliable.  So, for the measurements described in this paper, it was critical to find an operating regime in which the plasma and dust particles could remain stable and without distortion over the entire range of magnetic field settings. 
%%%%%%%%%%%%%%%%%%%%%%%
%%%%%%%%%%%%%%%%%%%%%%%%
%
\section{Result and discussion}\label{sec:results}
\begin{figure}[!htbp]
\includegraphics[scale=0.80]{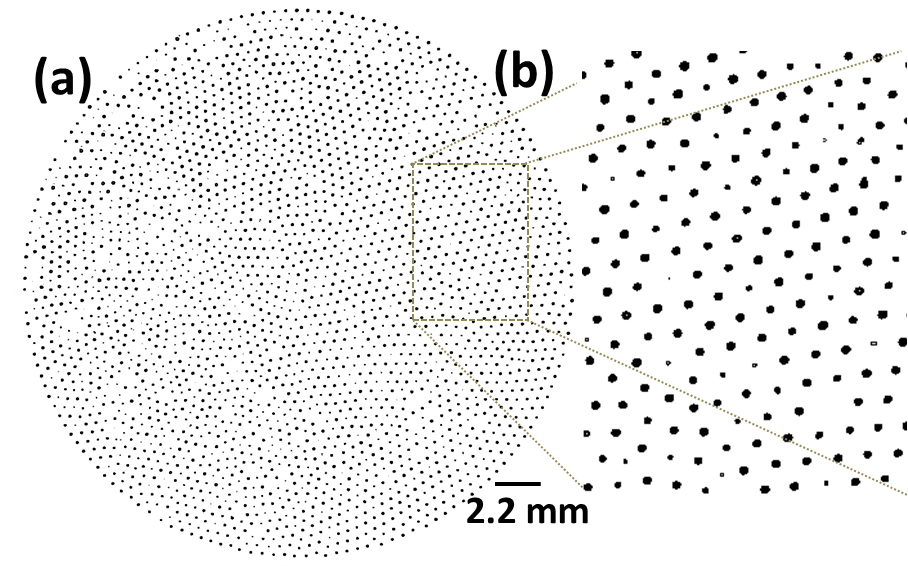}
\caption{\label{fig:crystal} The real picture of a) coulomb crystal and its b) zoomed view, formed at B = 0, P = 221 mTorr, rf power = 3.5 Watt.}
\end{figure}

\begin{figure}[!htbp]
\includegraphics[scale=0.55]{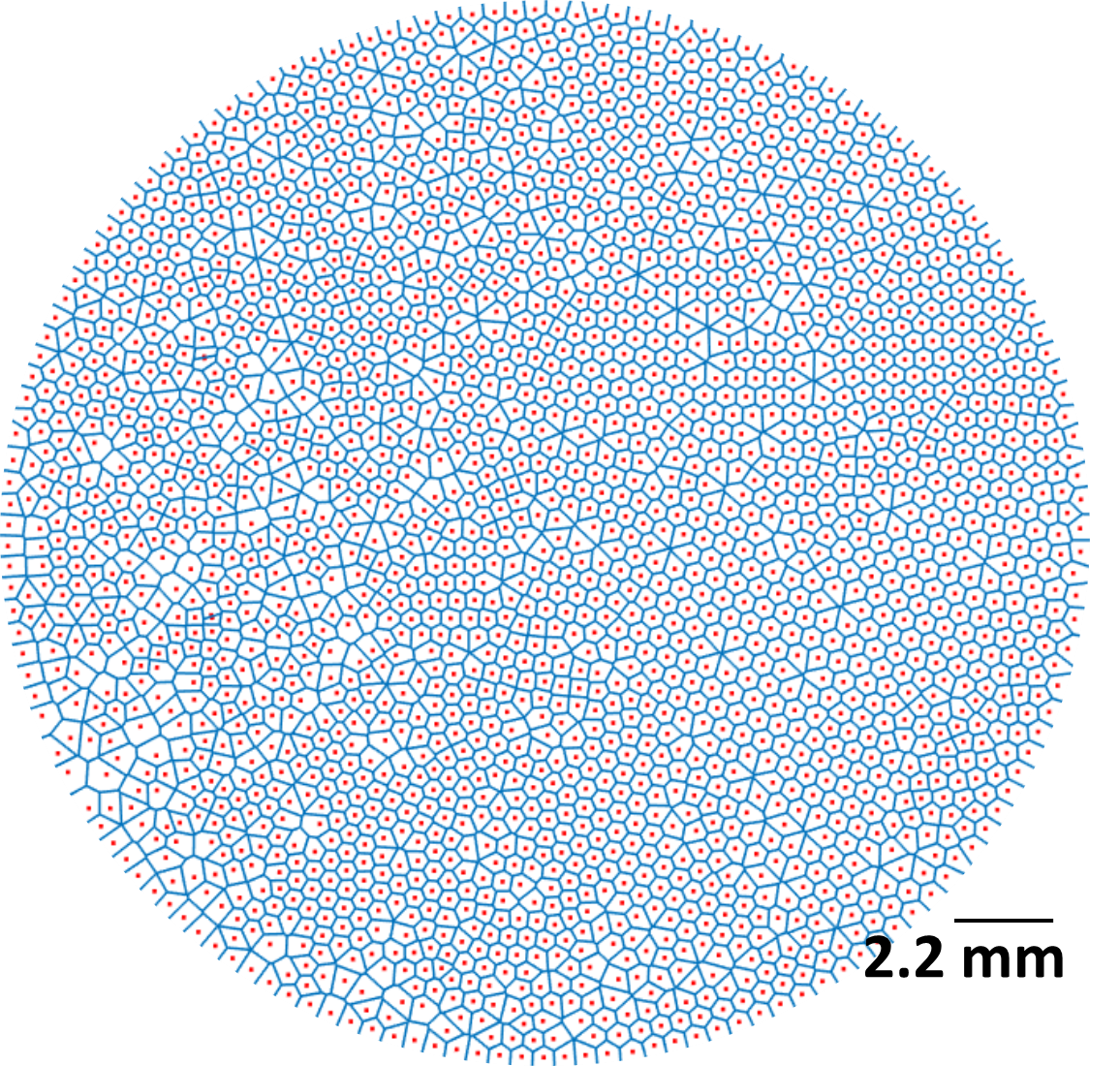}
\caption{\label{fig:voronoi_crystal} Voronoi diagram of the particles location at B = 0 which is shown in Fig.~\ref{fig:crystal}.}
\end{figure}

\begin{figure}[!htbp]
\centering
\includegraphics[scale=0.5]{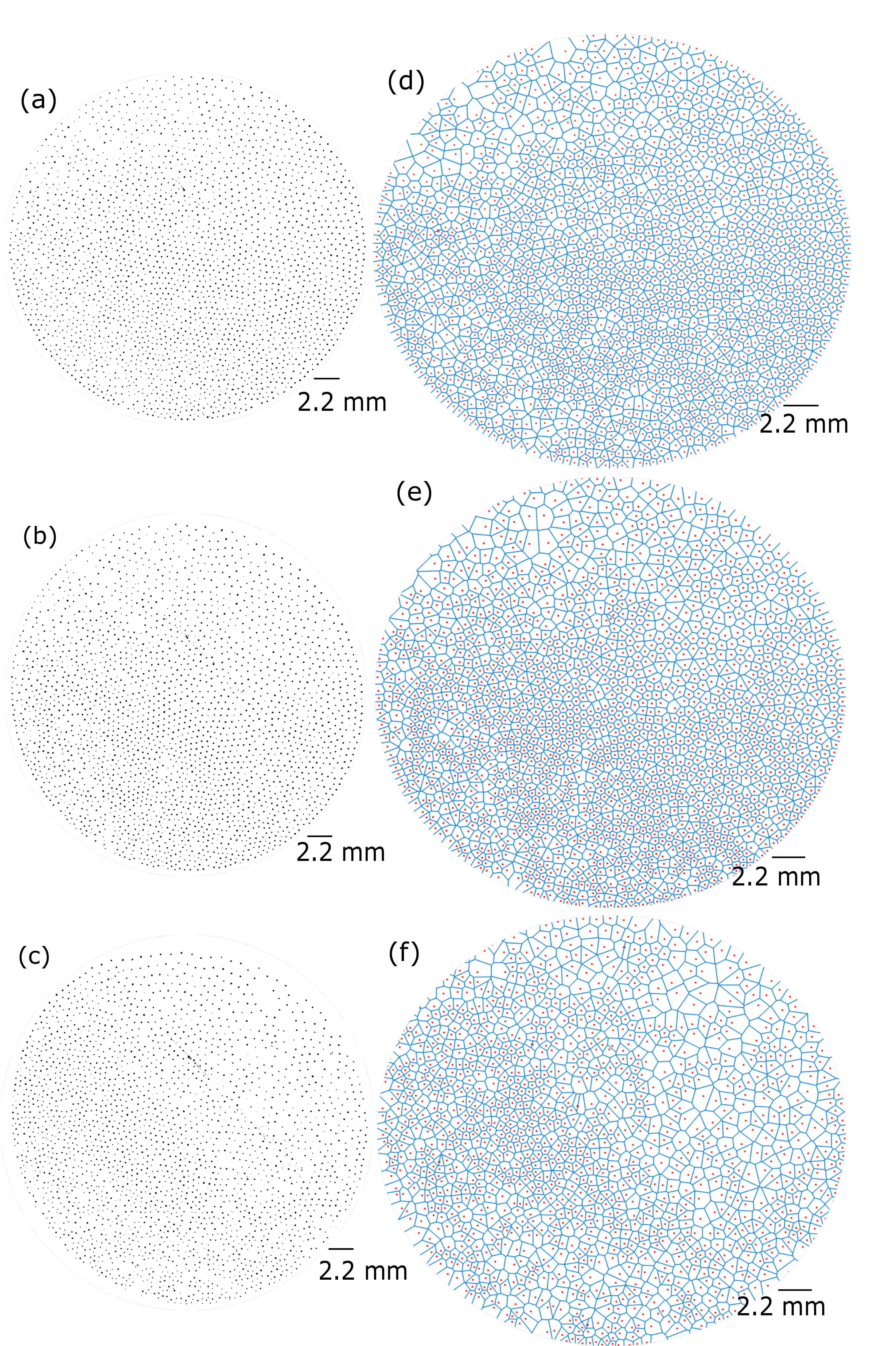}
\caption{\label{fig:voronoi_crystal_magnetic_field} Inverted image of Particles configurations and corresponding Voronoi diagram with changing magnetic field; a) and d) corresponds the structural configuration at B = 0.512 T, b) and f) for B = 0.896 T and, c) and f) are the inverted snapshot and corresponding Voronoi plot at B = 1 T, delineating the liquid state.}
\end{figure}

%
%%%%%%%%%%%%%%%%%%%%%%%%%%%%%%%%%%%%%%%%%%%%%%%%%%%%%%%%%%%%%%%%%%%%%%%%%%%%%%%%%%%%%%%%%%%%%%%%%%%%%%%%%%%%%%%%%%%%%%%%%%%%%%%%%%%
Based on visual observation, we have found that as we set the rf power of 3.5 Watt at $221\pm0.5$ mTorr ($29.4\pm0.1$ Pa) argon pressure, the particle cloud settled into a crystalline structure within the confining ring. For example, Fig.~\ref{fig:crystal} shows an inverted image of particles cloud forming a crystal. From the Voronoi diagram in Fig.~\ref{fig:voronoi_crystal}, it can be clearly seen that the particle cloud are arranged into a nearly hexagonal crystalline structure, although there is some disordering observed in the left region of the cloud. We believe that this may be due to formation of another layer below the first one. However, most of the crystal points do not show any deviation; this indicates that particles are almost aligned in vertical direction also. The structure are almost stationary with a small thermal fluctuation around the equilibrium position that we have checked from the overlapped image of 150 consecutive frames. We kept these discharge parameters constant while varying the external magnetic field. Fig~\ref{fig:voronoi_crystal_magnetic_field}(a) shows the inverted image of particle configuration at B$=0.512$ T, the corresponding Voronoi diagram is shown in Fig~\ref{fig:voronoi_crystal_magnetic_field}(d). It can be clearly seen that system ordering are getting changed and tending to disordered structure. It become even more disordered with further increase of the magnetic field (shown in Fig.~\ref{fig:voronoi_crystal_magnetic_field}(b) and associated Voronoi diagram in Fig.~\ref{fig:voronoi_crystal_magnetic_field}(e)) and achieve almost a new phase state when we reached the magnetic field of B$=1$ T. The Voronoi diagram of particle cloud at B$=1$ T (Fig.~\ref{fig:voronoi_crystal_magnetic_field}(f)) depicts the disordered structure which is nearly a liquid state.\par
%%%%%%%%%%%%%%%%%%%%%%%
%%%%%%%%%%%%%%%%%%%%%%%%
%%%%%%%%%%%%%%%%%%%%%%%
%%%%%%%%%%%%%%%%%%%%%%%%
%%%%%%%%%%%%%%%%%%%%%%%
\begin{figure}[!htbp]
\includegraphics[scale=1.0]{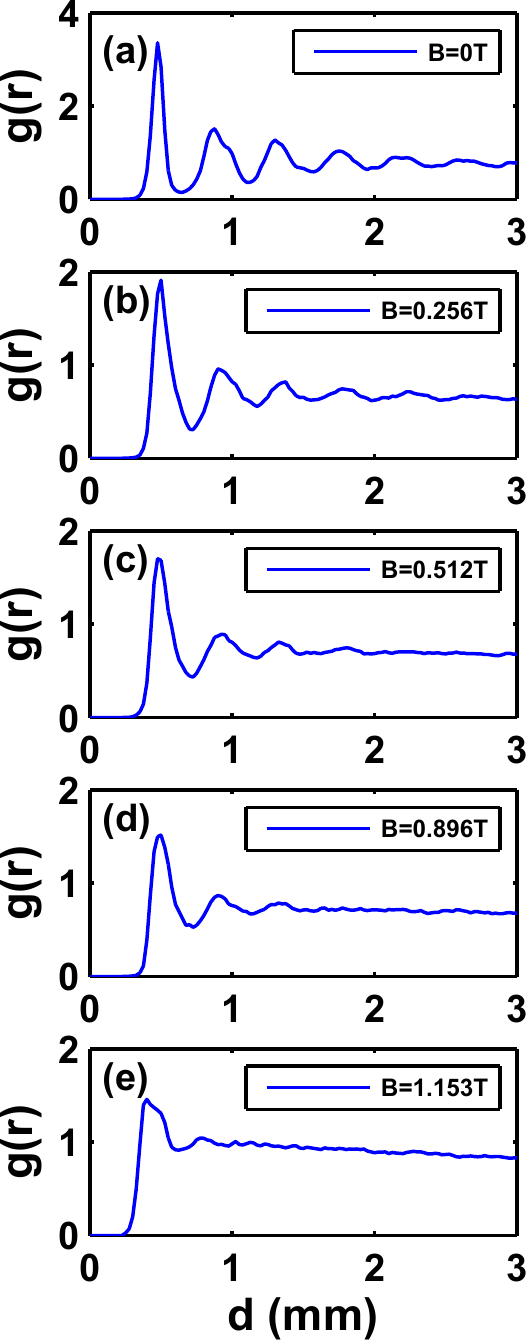}
\caption{\label{fig:gr_r} The pair correlation function g(r) of the particle clouds with changing magnetic field a) 0 T, b) 0.256 T, c) 0.512 T, d) 0.896 T, e) 1.153 T.}
\end{figure}

%%%%%%%%%%%%%%%%%%%%%%%%
For further verification of the change in the structure of the particle cloud with increasing magnetic field, we have plotted the pair correlation function, g(r) of the particle cloud in Fig.~\ref{fig:gr_r}. Here, g(r) is plotted for each magnetic field setting from B = 0 to 1.153 T. It has been calculated by directly measuring the average distance between particles where 250 frames have been used for averaging. It is important to be mentioned that the discharge parameter are kept fixed at the initial rf power of 3.5 Watt and initial pressure of 221 mTorr where we have observed perfect crystal shown in Fig~\ref{fig:crystal}. Fig.~\ref{fig:gr_r}(a) represents the g(r) vs interparticle distance (d) at zero magnetic field. As we can see from the figure, the nature of the correlation function shows the existence of long range ordering between the particles with a very pronounced peaks which is indicative of system being in crystalline state. With the increase of magnetic field the peaks become shorter and flattered, showing the increasing disorderness of the cloud, also the number of peaks are getting decreased. Fig.~\ref{fig:gr_r}(c) showing the pair correlation function at B$=0.512$ T, illustrate that the phase state of the particle changes significantly and they tends towards liquid state with primary peak followed by a fast descending second or third peak. We found that above 1 T of magnetic field the system become almost a liquid like that can be clearly seen in Fig.~\ref{fig:gr_r}(e), where only a small hump appears in correlation function.\par
% This observation is also supported by the plot of Fig. 2,where the variation of the long-range order with external magnetic field is shown.
%%%%%%%%%%%%%%%%%%%%%%%
%%%%%%%%%%%%%%%%%%%%%%%%
%%%%%%%%%%%%%%%%%%%%%%%
\begin{figure}[!htbp]
\includegraphics[scale=.60]{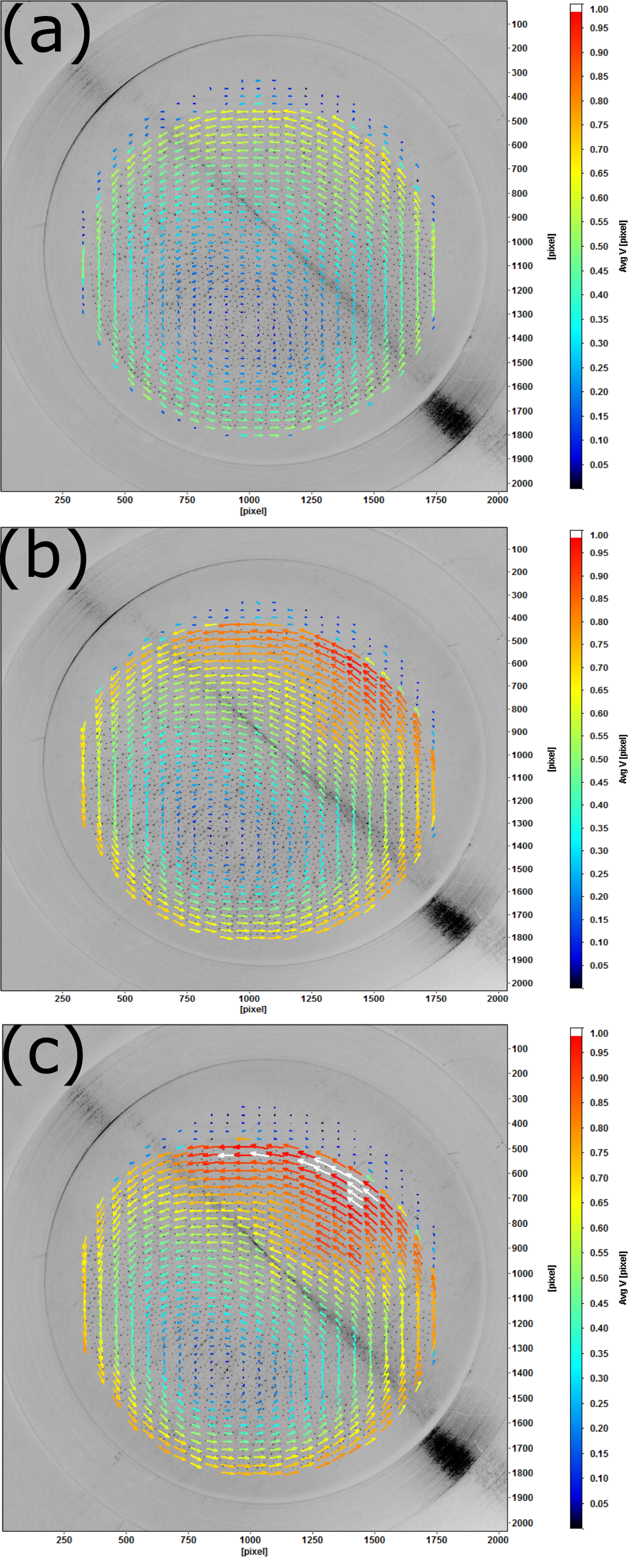}
\caption{\label{fig:velocity_map} Velocity vector field along with the magnitude of the velocity component v$_\theta$ at a) 0.512 T, d) 0.896 T and, e) 1 T.}
\end{figure}
\begin{figure}[!htbp]
\centering
\includegraphics[scale=0.8]{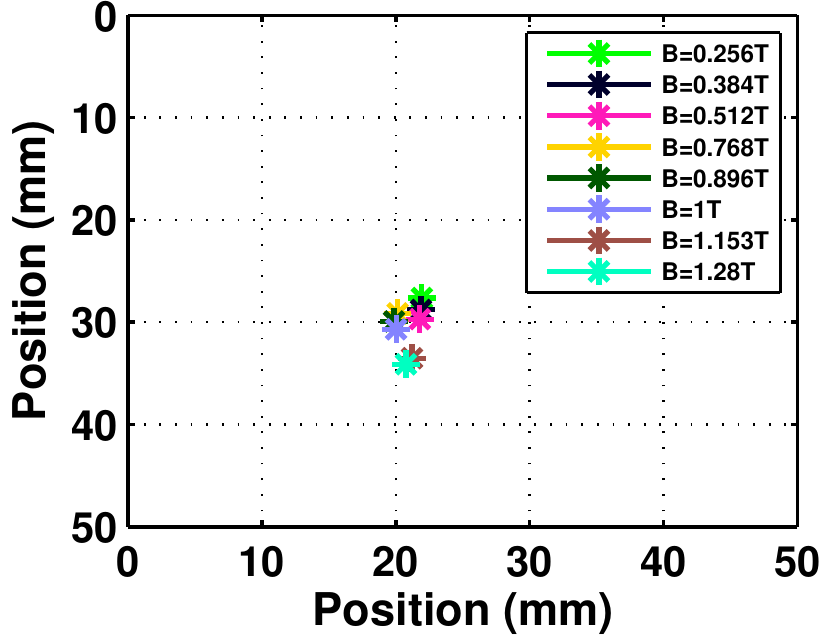}
\caption{\label{fig:rotation_shift} Shift in the center of rotation with the magnetic field from B = 0.256 - 1.28 T.}
\end{figure}

%%%%%%%%%%%%%%%%%%%%%%%%
In order to diagnose the cause of melting, we have inspected the other dynamical changes occurred in the particle dynamics in the influence of magnetic field. It is hitherto reported by several authors \cite{konopka_2000, uchida, sato, ed_5, ishihara} that adding a vertical magnetic field causes rotation of particle cloud as a whole due to an azimuthal E$\times$B ion drift. Therefore, we have also characterized the effective particle cloud rotation with changing magnetic field. Fig.~\ref{fig:velocity_map} shows the velocity vector field along with the magnitude of the azimuthal velocity component ($v_\theta$) of a particle cloud rotating in the presence of the magnetic field. Using the particle image velocimetry (PIV) analysis package, DAVIS 8 \cite{davis} tools are used to construct these velocity vector field. It is found that the rotation velocity increases with the increase of magnetic field. By performing the PIV technique it is easy to calculate the radial variation of angular velocity as a function of changing magnetic field. To do this, we have first determine the rotation center of velocity magnitude by fitting it with 2D elliptic contours. We observe a slight shift in the rotation center as we increase the magnetic field strength which represents that the crystals are not exactly symmetric about the center of confinement ring. However the shift is not much significant as can be seen from Fig.~\ref{fig:rotation_shift}.
%%%%%%%%%%%%%%%%%%%%%%%
%%%%%%%%%%%%%%%%%%%%%%%%
%%%%%%%%%%%%%%%%%%%%%%%%%%%%%%%%%%%%%%%%%%%%%%%%%%%%%%%
\begin{figure}[!htbp]
\centering
\includegraphics[scale=0.8]{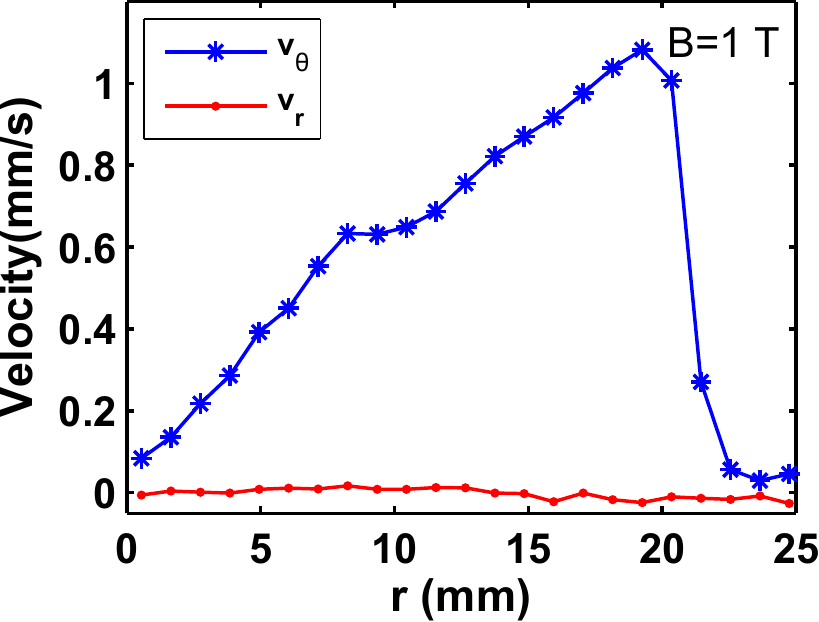}
\caption{\label{fig:vr_vt_comp}Comparison of radial and azimuthal velocity component for B = 1.0 T.}
\end{figure}
%%%%%%%%%%%%%%%%%%%%%%%%
After determining the rotation center, we first compare the velocity component in radial and azimuthal direction to get an idea about their contribution in the calculation of velocity magnitude of the particle cloud. Fig.~\ref{fig:vr_vt_comp} shows the $v_\theta$ and $v_r$ component for higher value of magnetic field (at 1 T). It can be seen from the figure that the $v_r$ contribution is very small and it is almost close to zero. This gives us a confidence that the the magnitude of the velocity is primarily the azimuthal velocity. We then calculate the angular velocity of each particle form the measured azimuthal displacement between consecutive frames and its distance from the rotation center. The data have been binned at the interval of 50 pixels (1 mm) in radial direction and averaged in each bin so as to reduce the uncertainties in the calculated values. \par
\begin{figure}[!htbp]
\centering
\includegraphics[scale=0.45]{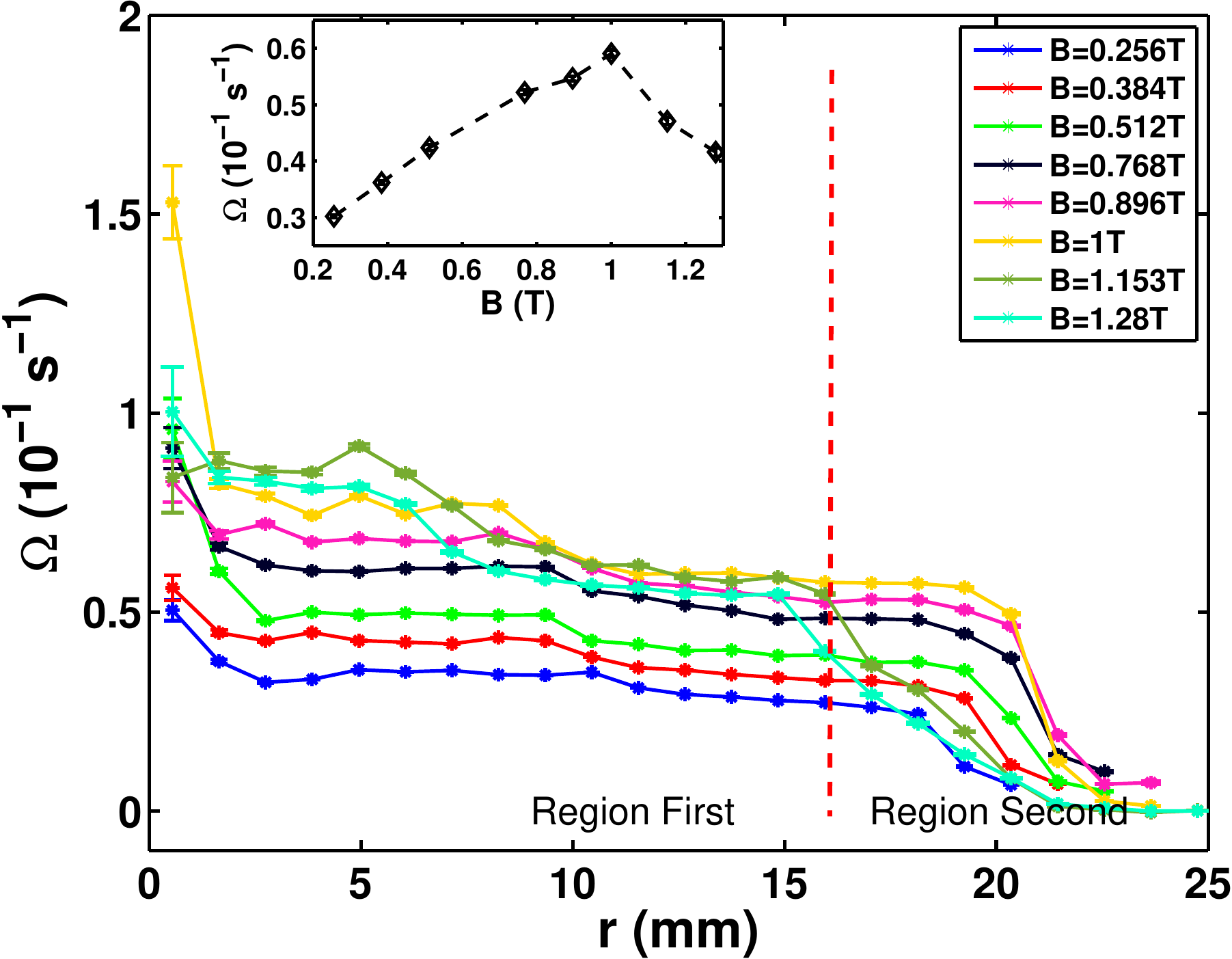}
\caption{\label{fig:omega_profile}Angular velocity $\Omega$ of the particles cloud rotating in the magnetic field vs distance from the center of rotation . The inset figure shows the variation of angular velocity with magnetic field from 0.256 to 1.28 T.}
\end{figure}

 The radial dependence of the particle angular velocity $\Omega$ with varying magnetic field from 0.256 to 1.28 T is shown in Fig~\ref{fig:omega_profile}. Two regions in the velocity profile has been observed, the inner region (region first) and profile edge region (region second). In inner region, the angular velocity curve do not show any significant radial dependence upto a magnetic field of 0.896 T, however a slight shift is observed at a position of 10 mm for all the magnetic field values. Above 1 T of magnetic field this shift become more prominent at the same position and construct a third region where the angular velocity decreases abruptly. Therefore, in a rough sense we can say that the curve shows three regions. In first region the angular velocity is almost constant upto 10 mm, then a slight decay in $\Omega$ is observed at 10 mm whereas, beyond this position the curve shows no any radial dependence upto 16 mm. In the edge region (beyond 16 mm), the angular velocity decreases with increasing position, exhibiting a strong radial dependence. Another interesting features illustrating the increase in angular velocity with the increase of magnetic field. As can be seen from the figure (inset), the angular velocity is increasing with increasing magnetic field upto 1 T. Above 1 T it is slightly decreasing  at 1.15 and 1.28 T.\par
%%%%%%%%%%%%%%%%%%%%%%%%%%%%%%%%%%%%%%%%%%%%%%%%%%%%%%%
%%%%%%%%%%%%%%%%%%%%%%%%
 Therefore, we interpret the melting of the plasma crystal with increasing magnetic field in the following manner. As the magnetic field is increased, the particle cloud shifted towards the confining ring.  The radial electric field near the ring leads to an azimuthal ion E $\times$ B drift, which couples to the dust grains and leads to their rotation as described by Uchida \textit{et al.} \cite{uchida, sato} and Ishihara and Sato \cite{ishihara}.  In this experiment, the measurements presented in Figs.~\ref{fig:vr_vt_comp} and \ref{fig:omega_profile} show that a differential rotation is established between the inner and outer regions of the plasma crystal.  This discontinuity in the angular velocity appears to provide the crystal with a breaking point that allows melting to occur with the increasing magnetic field.  Above B = 1 T, it is shown that the crystal can no longer retain its long-range order and the particles make the transition to a liquid-like state.
%%%%%%%%%%%%%%%%%%%%
\section{Summary}
In summary, this paper reports on an experimental observation of the melting of a plasma crystal as a function of magnetic field.  These experiments, performed using the Magnetized Dusty Plasma Experiment (MDPX) facility, studied the spatial ordering and structure of monodisperse melamine-formaldehyde dust particles that were held in a capacitively coupled, argon rf glow discharge at a fixed pressure and fixed rf power, while varying the magnetic field. The measurements show that at lower magnetic field (B $< 0.5$ T), there is an initial rigid rotation of the plasma crystal. With increasing magnetic field, the outer portion of the crystal develops a differential rotation that eventually leads to heating and melting of the crystal. It is noted that various theories have considered changes in the particle interaction potential may lead to crystal melting in a magnetic field. While these mechanisms may still play a contributing factor to the melting process, the experiments reported here show that the induced rotation is the dominant mechanism.
\label{sec:conclusion}
%\section*{References}
\section{Acknowledgements}
The authors would like to thank Prof. Uwe Konopka for providing us with a source of monodisperse microparticles and his advice when carrying out these experiments.  This work is supported by grants from the National Science Foundation, PHY-1613087, and the US Department of Energy, DE - SC0016330 and DE - SC0010485.
%###################################################################################

\end{document}